\documentclass[]{aa}

\usepackage{graphicx}
\usepackage{txfonts}

\begin{document}

   \title{The BAST algorithm for transit detection}

   \author{S. Renner \inst{1}, H. Rauer \inst{1,2}, A. Erikson \inst{1}, P. Hedelt \inst{1},  
           P. Kabath \inst{1}, R. Titz \inst{1} \and H. Voss \inst{3}}

   \offprints{S. Renner, \\ \email{Stefan.Renner@dlr.de}}

   \institute{Institut f\"ur Planetenforschung, Deutsches Zentrum f\"ur Luft-
   	      und Raumfahrt (DLR),
              Rutherfordstra$\beta$e 2, 12489 Berlin, Germany
	   \and
	      Zentrum f\"ur Astronomie und Astrophysik, Technische Universit\"at, Hardenbergstr. 36, 10623 Berlin, Germany
	   \and
	      Departament d'Astronomia i Meteorologia, Facultat de F\'isica, Universitat de Barcelona, 
	      Barcelona, C/ Mart\'i Franqu\'es 1, 08028 Barcelona, Spain}

   \date{Received / Accepted 26 September 2008}


\abstract
   {The pioneer space mission for photometric exoplanet searches, CoRoT, steadily monitors about 
   12000 stars in each of its fields of view. Transit detection algorithms are applied to derive promising 
   planetary candidates, which are then followed-up with ground-based observations.}
   {We present BAST (Berlin Automatic Search for Transits), a new algorithm for periodic transit detection, 
   and test it on simulated CoRoT data.}
   {BAST searches for box-shaped signals in normalized, filtered, variability-fitted, and unfolded light 
   curves. A low-pass filter is applied to remove high-frequency signals, and linear fits to subsections
   of data are subtracted to remove the star's variability. A search for periodicity is then performed 
   in transit events identified above a given detection threshold. Some criteria are defined to better 
   separate planet candidates from binary stars.}
   {From the analysis of simulated CoRoT light curves, we show that the BAST 
   detection performance is similar to that of the Box-fitting Least-Square (BLS) method if the signal-to-noise ratio is 
   high. However, the BAST box search for transits computes 10 times faster than the BLS method.
   By adding periodic transits to simulated CoRoT data, we show that the minimum periodic depth detectable with BAST
   is a linearly increasing function of the noise level. For low-noise light curves, the detection limit corresponds to a 
   transit depth $d \simeq 0.01$ \%, i.e. a planet of 1 Earth radius around a solar-type star.}
   {}


   \authorrunning{S. Renner et al.}
   \maketitle


\section{Introduction}

   The CoRoT satellite (Baglin et al. \cite{baglin06}), successfully launched in 
   December 2006, is the first space-based photometric experiment aimed at 
   discovering extrasolar planets. With a very accurate photometric signal 
   and a continuous time sampling over long periods, it should have
   the capability of detecting extrasolar planets with sizes down to a couple 
   of Earth radii (Bord\'e et al. \cite{borde03}), and will
   observe a total of 60000 stars in 5 long runs of 150 days each. The first 
   of these long observing runs was performed between May and October 2007, 
   and the scientific analysis of the data is now ongoing. The first detections 
   of transiting planets observed by CoRoT have already been reported (Barge et al.
   \cite{barge08}, Alonso et al. \cite{alonso08}, Bouchy et al. \cite{bouchy08}, 
   Aigrain et al. \cite{aigrain08}, Moutou et al. \cite{moutou08}).
   
   Transit-detection algorithms are used to derive planetary candidates from
   the analysis of the stellar light curves. Several methods were recently tested 
   with simulated CoRoT light curves (Moutou et al. \cite{moutou05}, \cite{moutou07}). 
   For instance, image processing techniques (Guis \& Barge \cite{guis05}),
   removals of low-frequency harmonics, and iterative non linear filters 
   (Aigrain \& Irwin \cite{aigrain04}) were used to remove the stellar variability. 
   The transit-search methods included a simple correlation with a sliding transit template, 
   matched filters (Jenkins et al. \cite{jenkins96}; Bord\' e et al. \cite{borde07}), and several
   box-shaped search methods such as the Box-fitting Least-Square (BLS) 
   algorithm (Kov\'acs et al. \cite{kovacs02}). A theoretical comparison of these
   detection methods was proposed (Tingley \cite{tingley03a}, \cite{tingley03b}), 
   which concluded that no detector is clearly superior for all transit signal
   energies, but an optimized BLS algorithm still performs slightly better for
   shallower transits.
   
   In this paper we present BAST (Berlin Automatic Search for Transits), a new algorithm
   for periodic transit detection. Using a subset of simulated CoRoT data from Moutou et al. 
   (\cite{moutou05}, \cite{moutou07}), we compare
   its performance with another box-shaped search technique, the  
   BLS algorithm (Kov\'acs et al. \cite{kovacs02}). By adding periodic transit events to the
   light curves, we assess the BAST detection capability with respect to the increase in the 
   noise level.


\section{The BAST algorithm}

Basically, BAST searches for box-shaped signals in normalized, filtered,
variability-fitted, and unfolded light curves. It is designed to detect
periodic transit events and is based on the preliminary search routine for transits
used by the team at DLR (German Aerospace Center, Berlin) in the first blind-test analysis 
(Moutou et al. \cite{moutou05}).
BAST, a sun goddess in Egyptian mythology, is here an acronym for 
{\bf B}erlin {\bf A}utomatic {\bf S}earch for {\bf T}ransits. 
In the following sections, we describe the successive steps, illustrated in Fig.\ref{bast_algo}, 
of the BAST algorithm. The parameters used are adapted
to the CoRoT mission, for which the temporal sampling of the light curves is 512 s
with a total duration of 150 days. Nevertheless, the algorithm 
could easily be applied to future space-based transit detection surveys like Kepler or
PLATO.

\subsection{Light-curve filtering}
\label{filtering_section}

First, the light curves are normalized (Fig.~\ref{bast_algo}a).
Then, a standard low-pass filter is used to eliminate high-frequency signals (Fig.~\ref{bast_algo}b). The filtering
is performed in the frequency domain using the fast Fourier transform. All the frequencies
higher than a given cut-off value are removed from the original signal. The
dominant periodic signal in the CoRoT light curves is at a period of $\simeq 1.68$ h, 
corresponding to the orbital period. The cut-off period is then varied between $\simeq 1.5$  
and 7 h for the filtering, with a step size of $\simeq 0.3$ h.
Since one does not expect to detect transiting planets with periods of only a few hours, 
this value of 7 h is acceptable. The variation of the cut-off period allows the high-frequency noise
to be removed at different levels. 

The shape of the transit signals is 
moderately deformed by this kind of filtering, but the influence of this effect is negligible 
for the purpose of a detection tool. Another side-effect of the low-pass filtering is that an 
additional modulation of the light curves occurs at the beginning and the end of the
data. Therefore the first and last few days of data (typically between 
1 and 10 days) are excluded from the transit search.

\subsection{Stellar variability subtraction} 
\label{fit_section}

\begin{figure}
   \centering
   \includegraphics[width=8.5cm]{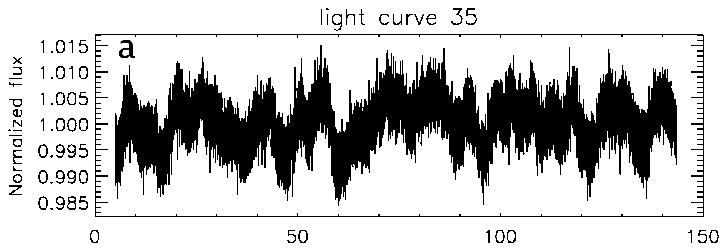}
   \includegraphics[width=8.5cm]{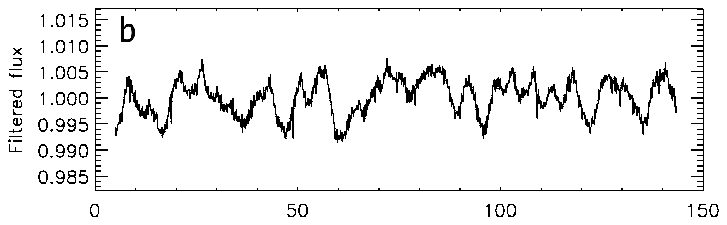}
   \includegraphics[width=8.5cm]{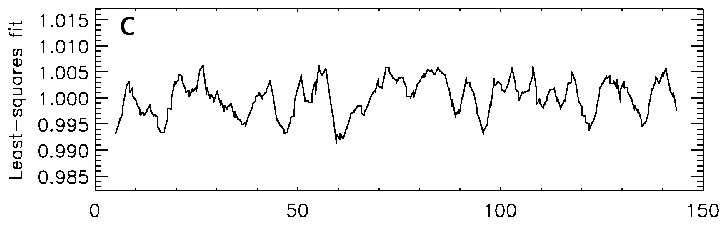}
   \includegraphics[width=8.5cm]{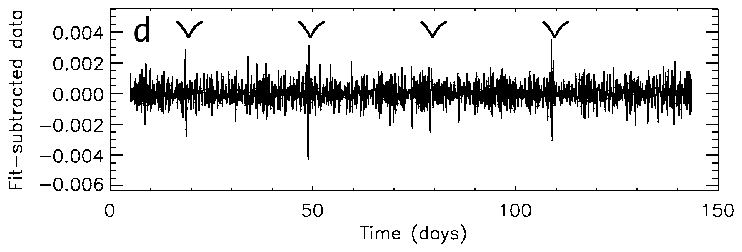}
   \includegraphics[width=8.5cm]{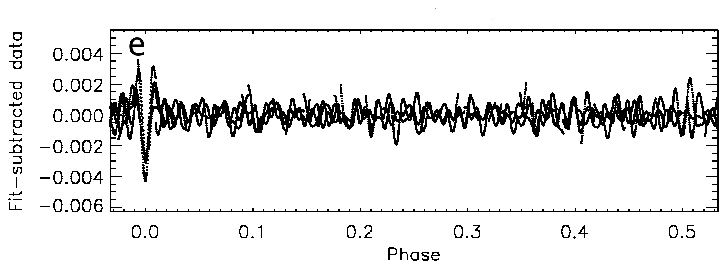}
      \caption{The successive steps in the BAST algorithm, illustrated with the analysis of 
      a white light curve from the second CoRoT blind test (Moutou et al. \cite{moutou07}). 
      The target is number 35 in the database.
      From top to bottom: (a) the light curve is normalized; (b) a low-pass filter is 
      applied to remove high-frequency signals; (c) the stellar variability is modeled by least-squares
      fitting to subsections of data; (d) the fit is subtracted from the filtered light curve, 
      and a box search for periodic transits is performed. The periodic signal is indicated with 
      arrows on the plot; (e) the final low-pass filtered, variability-fitted light curve is
      phase-folded.}
         \label{bast_algo}
   \end{figure}

The stellar variability is fitted locally (Fig.~\ref{bast_algo}c). The light curves are separated into
subsections of equal lengths and a linear least-squares fit to the data is performed 
for every subsection. The number of subsections is varied from 250 to 20, with a step 
size of 5, to model the variability at different scales. 
For a light curve of 150 days, the size of the subsections is then 
between $\simeq 0.5$ and 7.5 days. This range of values ensures that no transit 
events are significantly altered. 
In a subsequent step, the fit is subtracted from the data (Fig.~\ref{bast_algo}d).

\subsection{Box search for transits}
\label{search_section}

We compute the standard deviation $\sigma$ of the normalized, low-pass filtered, and 
variability-fitted light curves. Then a box search for transit-like events is carried out. 
All data points deviating from the average signal by $k \times \sigma$ are identified, where $k$ is a 
variable parameter starting at a value of $\simeq 2$. The neighboring deviating points are combined 
into one single detection. More precisely, consecutive dips separated by less than 
$\simeq 20$ h are replaced by their mean value. Dips corresponding to transit lengths well
below 1 hour are excluded. The resulting mean epochs are listed.
Thereafter, the times of the transit-like events found are automatically searched for 
periodicity: time differences between all detected events are estimated and retained 
when a single time difference has occurred more than twice within a given error margin. 
This margin is fixed at 0.1 days. The variable parameter $k \geq 2$ used to 
identify deviating points increases by values of 0.1 until a 
periodic signal (or no dips at all) is found.  

We point out that this method should allow us to also detect transits when the transit times vary,
e.g. due to gravitationally interacting planets.

\subsection{Criteria for identifying planetary candidates}
\label{criteria_section}

Once the period has been determined, we estimate the depth and the duration of the transits
from the low-pass filtered and variability-fitted light curves. 
The depth is defined as the mean of the minimum values of all the transit events.
For each dip, we estimate the center and the length of the transit.  
The duration is then defined as the mean of all the segments determined. 

Following Seager \& Mall\'en-Ornelas (\cite{seager03}) and using Kepler's third law, the transit 
duration $t_T$ for a circular orbit with an impact parameter $b=0$ is given by
\begin{equation}
t_T \simeq \frac{P R_*}{\pi a} = 1.8 R_* \Big{(} \frac{P}{M_*} \Big{)}^{1/3},
\label{tr_dur}
\end{equation}

\noindent where  $t_T$ is in hours, $R_*$ and $M_*$ are respectively the radius and the mass of the star
in solar units, $P$ is the planet's period in days, and $a$ its semi-major axis. 
We compute $K=t_T / P^{1/3}$, and use the star information to flag transit candidates for which 
K is much greater than $1.8 R_* / M_*^{1/3}$. Too long a measured transit duration could indeed 
be due to a problem with the transit parameters, such as an incorrect period. It could also
arise from the star parameters, hinting at the possibility of a background eclipsing 
binary with a spectral type different than that of the target. However, this criterion must be used
with care, since there are eccentric planets (even at short periods), which can lead to longer transit durations.
On the other hand, candidates with very short transit durations (below 1 hour) are rejected.
A condition on the transit depth ($d>5-10$ \%) is also set to identify obvious eclipsing binaries. 

A strong indicator for distinguishing planets from binary stars is the presence of a
secondary eclipse in the light curve. Therefore, the phase-folded light curves of the transiting 
candidates are scanned for shallower transit events, in particular at half of the candidate's period.
A phase-folding at twice of the period derived is also performed to detect possible small 
depth differences.

Modulations can appear in the baseline of the light curves at twice the orbital frequency, 
which are caused by the tidal forces exerted by a massive stellar companion. Moreover, V-shape transits 
indicate binaries, too. These phenomena (ellipsoidal variability and V-shape transits) are visually 
inspected. Because grazing planetary transits also produce V-shape signals, let us note that candidates 
should not be rejected based on the light curve shape alone. 

A significant part of the CoRoT targets are observed in three spatially-separated colors 
(red, green, blue, but not corresponding to defined photometric systems).
Because only planets produce the same transits in the three channels, the transit 
chromaticity can also be an indicator of binaries. However, the rejection of planet candidates based on 
the R, G, B colors must be made with caution, since a background star in one of the channels 
may dilute the planetary transit and then mimic a chromatic effect. 
We do not apply automatic chromatic criteria. We instead choose to visually check candidates in the 
3 color channels, and identify those showing a certain level of chromaticity.

\subsection{Summary of BAST}

To sum up, the BAST algorithm runs as follows: 
\begin{itemize}
\item the filtering and the fitting of the stellar variability are applied to each light curve 
(Sects. \ref{filtering_section} and \ref{fit_section}), exploring a range of  parameter values 
(cut-off period and size of the fitted subsections).
\item A transit search is made (Sect. \ref{search_section})
for each cut-off period value and each size of the fitted subsections, and periodic transits found
above the detection threshold are stored.
\item The transit parameters (period, depth, duration) are calculated, and the possible different 
detected periods are sorted on the period values.
\item Automatic checks are then performed for the transit duration, transit depth, 
depth differences, and the presence of a secondary eclipse (Sect. \ref{criteria_section}). 
If several periodic signals remain (such as period multiples), the correct one is then easily identified by eye.
\item Finally, the shape of the transits and possible ellipsoidal modulations are visually checked.
\end{itemize}

A detailed example is given in Fig.~\ref{bast_algo}, with the analysis of a simulated CoRoT white light 
curve (number 35) from the 
second CoRoT blind test exercise (Moutou et al. \cite{moutou07}). A transiting planet of 1 Jupiter radius in 
orbit with a period of 30.1 days around an A5V star (corresponding to a depth of 0.3 \%) was added to the 
light curve. The different steps in the BAST algorithm are displayed, the periodic transits are 
found, and the transit parameters are retrieved. 
The algorithm parameters used for this transit detection were: 
a cut-off period value of 6.25 h, a size of the subsections of 0.9 d, and $3.4 \sigma$
to identify dips deviating from the average signal. The parameters retrieved from the filtered,
fit-subtracted light curve are: period $P=30.11$ d, depth $d=0.35$ \%, and duration $t_T=5.2$ h.
On the phase-folded light curve, we can notice again  
that the shape of the transits is slightly deformed by the filtering. A more detailed analysis of the
original light curves can provide accurate transit parameters once a candidate is found. 
However, this is beyond the scope of this paper.
An interesting related work that reconstructs a transit signal in the presence of stellar variability 
has recently been proposed (Alapini \& Aigrain \cite{alapini08}).


\section{Application to simulated CoRoT light curves}

Two blind detection tests were conducted before the arrival of CoRoT data. In both cases, a set of light 
curves was generated, including noise sources, stellar variability models (Lanza et al. 
\cite{lanza04}; Aigrain et al. \cite{aigrain04a}), and a model of the CoRoT
instrument. Then, simulated planetary transit events and a range of contaminating events
that mimic transits (eclipsing binaries, variable stars) were added to the light curves.  

The first blind test (BT1) aimed at a comparison of the light curve analysis methods to assess the detection performance.  
A large sample of 1000 light curves was generated with few added transit-like events.
The results were published in Moutou et al. (\cite{moutou05}).

The primary goal of the second exercise (BT2) was to test the ability to characterize detected 
transits, and to distinguish transiting planets from eclipsing binaries, using the information
from the 3 color channels of CoRoT (Moutou et al. \cite{moutou07}). Thus, 237 light curves in red, green, 
and blue were built, and all of them contained transits either from planets or from binaries. 

Although no detector is clearly superior, the BT1 allowed to conclude that the most sensitive analysis 
is performed with periodic box-shaped detection algorithms, such as the BLS algorithm 
(Kov\'acs et al. \cite{kovacs02}). This result also appears in a theoretical comparison 
proposed by Tingley (\cite{tingley03a}, \cite{tingley03b}). Different approaches to 
analyse the light curves are nevertheless important, since it was also shown that the generation of 
false positives was method-dependent.

\subsection{BAST applied to BT2 data}
\label{bast_bt2} 

Here, we search transits in the BT2 white light curves with the BAST algorithm. 
We only use the first 100 light curves (ID 0-99) that are produced assuming a good correction of the residual
light scattered from the Earth. For the real CoRoT data, this source of noise is corrected with the 
reduction pipeline.

We compare the results by also running
the BLS algorithm with the same filtering and detrending techniques as BAST 
(Sects. \ref{filtering_section} and \ref{fit_section}).
The BLS parameters used are the
following (Kov\'acs et al. \cite{kovacs02}): $n_f=6000$ frequency points in which the spectrum
is computed, $m=200$ bins to divide the folded time series, $f_{min}=0.01$ and $f_{max}=1$
day$^{-1}$ for the minimum and maximum frequencies, $q_{min}=0.01$ and $q_{max}=0.05$
for the minimum and maximum fractional transit lengths.

   \begin{figure}[t]
   \centering
   \includegraphics[width=8.5cm]{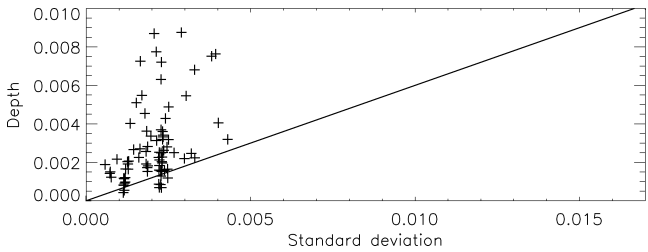}
      \caption{Depth versus standard deviation for the BT2 light curves 0-99. The line is the BAST 
      detection limit, see Sect.~\ref{det_capab}.}
         \label{depth_sigma_bt2}
   \end{figure}

\begin{table}
\caption{BT2 results from the analysis of the (white) light curves 0-99.}             
\label{blindres}      
\centering          
\begin{tabular}{c|c|c}         
     & BAST & BLS \\ 
\hline \hline                    
   correct $P$ or $P/2$ or $2P$ & 76 & 44 \\ 
\hline
    $P/2$ or $2P$  & 9 & 40 \\ 
\hline 
   wrong period & 12 & 16   \\
\hline 
   no detection & 3 & 0     \\
\hline 
   planets identified & 4 & 4     \\          
\end{tabular}
\end{table}

The light curves 0-99 of the BT2 all contain a transit event, and 12 planets from 0.1
to 1 Jupiter radius are included in 10 of them. The results of the transit search with BAST and
with the BLS algorithm are given in Table~\ref{blindres}. Both with the BAST and BLS methods,
85 \% of the detections are correct, or else affected by a factor of 2 in the period estimation
($P/2$ or $2P$), with respect to the nominal period of the simulated system. 
More precisely, BAST identifies 76 correct periods, and 9 detections are wrong by 
a factor of 2. The BLS method only finds 44 correct periods, and 40 are affected by a factor of 2 
with respect to the real period.
The eclipsing binaries 
with similar transits are probably responsible for such confusions. 

Note that we have also performed the BT1 exercise again with BAST and BLS, and we get similar results to
those in Moutou et al. (\cite{moutou05}).

The similar performances obtained here for BAST and BLS stem from the noise level of the BT2 
light curves being relatively low. The standard deviation range of the normalized light curves 
is $0.00057 \leq \sigma_{LC} \leq 0.00430$ for transit depths $d \leq 1$ \%, see Fig.~\ref{depth_sigma_bt2}. 
However, the box transit search of BAST is approximately 10 times faster 
than the BLS method (with the parameters given in the previous section).

\subsection{Detection capability versus noise}
\label{det_capab}

Here we show how the BAST algorithm behaves with increasing noise level. 
We selected a subset of $250$ BT1 light curves (not containing transit events) with various S/N; 
more precisely, the
standard deviation range of the normalized light curves of this subset is 
$0.00039 \leq \sigma_{LC} \leq 0.01640$. We added simulated periodic transit events to the data, with a 
fixed duration $t_T=3.5$ h and a fixed period $P=3.689857$ d.
Then, we slowly decreased the depth to derive the minimum detected signal for given standard deviations.
Changing the duration and period values would yield the same results 
(see the method to identify periodic transits in Sect.~\ref{search_section}).
For few cases, we also performed the same exercise using the BLS algorithm, with the same 
filtering/detrending techniques as BAST and the same parameters as in Sect. \ref{bast_bt2}.

\begin{figure}[t]
   \centering
   \includegraphics[width=8.5cm]{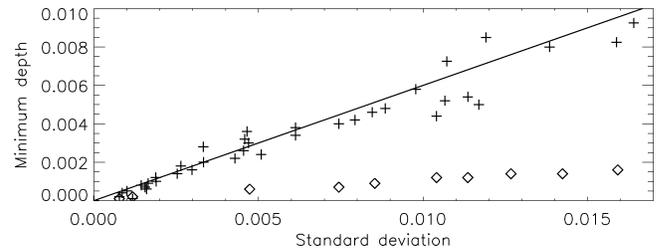}
      \caption{Minimum depth versus light-curve standard deviation. The plus and diamond signs correspond to BT1 light
      curves analyzed with BAST and BLS, respectively. A linear fit of slope $0.6$ has been included.}
         \label{bast_sensitiv}
   \end{figure}

We find that the BAST detection limit is a linearly increasing function of the noise level,
with a slope of 0.6. 
The results are shown in Fig.~\ref{bast_sensitiv}. For each point in the figure, the standard 
deviation given is an average based on the values of a greater number of light curves, 
all with the same minimum depths. This was done to give a clear overview of the general trend.
We notice that the BLS method is clearly more sensitive than BAST as the noise increases.
Nevertheless, BAST can detect periodic signals down to a transit depth $d=0.01$ \% (i.e. Earth-type planets) for low 
standard deviation values less than $\sigma_{LC} \sim 0.001$. Moreover, Fig.~\ref{bast_sensitiv} shows that 
comparable or smaller planets than Neptune ($d \leq \sim 0.3$ \%, depending on the type of the star) 
are detectable with BAST if $\sigma_{LC} \leq \sim 0.005$. About 40 \% of the monochromatic light curves 
of the first CoRoT observing run (IRa01) are within this limit.

\section{Conclusions and prospects}

We have developed BAST (Berlin Automatic Search for Transits), a box-shaped search algorithm to 
detect transits and identify planetary candidates. From the analysis of BT2
light curves (with relatively high S/N), we have shown that BAST and the BLS transit 
search method, combined with the same 
low-pass filtering and linear-fit detrending, perform similarly. The BAST transit search is  
10 times faster than the BLS method; indeed, transits are simply identified as individual events deviating 
from the average signal and then checked for periodicity. By adding simulated periodic transit events to BT1
light curves, we showed that the BAST detection limit is a linearly increasing function 
of the light curve noise level. Nevertheless,
BAST can detect small transiting planets down to the size of the Earth if the S/N is sufficiently high.
A detailed analysis of the detection capabilities of the various algorithms using real CoRoT data 
will be published in a future article. Such a study should take the removal 
of systematic errors into account (Tamuz et al. \cite{tamuz05}, Kov\'acs et al. \cite{kovacs05}).
More work is needed to include these corrections in BAST, which would in turn help 
to push the detection limit further.
Finally, we note that BAST should be sensitive to transiting planets with varying transit times (due to 
additional perturbing bodies), since it is working
on unfolded data and includes a tolerance for period variations.


\end{document}